\documentclass[aps,pra,twocolumn,floatfix,nofootinbib]{revtex4}

\usepackage{graphicx}
\usepackage{amsmath,amssymb}
\usepackage[T1]{fontenc}
\usepackage{mathptmx}
\usepackage[dvipsnames]{color}
\usepackage[a4paper=true,colorlinks=true,citecolor=myblue,linkcolor=black,urlcolor=myblue]{hyperref}
\definecolor{myblue}{named}{MidnightBlue}

\newcommand{\ket}[1]{|#1\rangle}
\newcommand{\bra}[1]{\langle#1|}
\newcommand{\be}{\begin{equation}}
\newcommand{\ee}{\end{equation}}
\newcommand{\ba}{\begin{eqnarray}}
\newcommand{\ea}{\end{eqnarray}}

\def\la{\langle}
\def\ra{\rangle}
\def\tr{\mbox{Tr}}
\def\unity{\mathbb I}

\begin{document}

\title{Geometric derivation of the quantum speed limit}

\author{Philip J. Jones}
\author{Pieter Kok}\email{p.kok@sheffield.ac.uk}

\affiliation{Department of Physics and Astronomy, University of Sheffield, Hicks building, Hounsfield Road, Sheffield S3 7RH, United Kingdom.}

\begin{abstract}
 \noindent The Mandelstam-Tamm and Margolus-Levitin inequalities play an important role in the study of quantum mechanical processes in Nature, since they provide general limits on the speed of dynamical evolution. However, to date there has been only one derivation of the Margolus-Levitin inequality. In this paper, alternative geometric derivations for both inequalities are obtained from the statistical distance between quantum states. The inequalities are shown to hold for unitary evolution of pure and mixed states, and a counterexample to the inequalities is given for evolution described by completely positive trace-preserving maps. The counterexample show that there is no quantum speed limit for non-unitary evolution.
\end{abstract}

\maketitle

\section{Introduction}

\noindent
In experiments, we often do not have direct access to the parameter we want to measure, but only to certain observable variables. We tend to formulate a model of the system that tells us how we expect the observable variables depend on the parameters of interest of the system. Given an estimator function based on these observables, we can then calculate the most likely value of the parameter based on the measurement data \cite{helstrom76,holevo82,holevo01,jaynes03}. When estimating an unknown parameter we will always be restricted by statistical uncertainty: the more tests we perform the greater our confidence in the result. More quantitatively, given a large number of independent tests $N$, the average error is at best $\Delta \theta = \sigma/\sqrt{N}$, where $\sigma^{2}$ is the variance in each test. In quantum metrology it is possible to beat this limit by using entanglement. The fundamental Heisenberg limit then allows us to estimate the parameter $\theta$ with average error $\Delta \theta \propto 1/N$, where the $N$ tests are no longer independent of each other. In order to accommodate entanglement across the $N$ tests, the estimation procedure must essentially boil down to a single-shot measurement. To distinguish between two configurations in a single-shot measurement with unit probability, the two configurations must be described by orthogonal quantum states \cite{caves81,yurke86,pezze08}. If $\theta$ parametrizes the evolution of the initial state to the final state, maximum sensitivity is obtained when the state of the system evolves to an orthogonal state for the smallest possible value of $\theta$. This is the geometric interpretation of parameter estimation \cite{wootters81,braunstein94,braunstein96,giovannetti06}.

The minimum value of $\theta$ needed for a quantum system to naturally evolve to an orthogonal state is determined by the moments of the generator ${K}$ of rotations (or translations) in $\theta$. In particular there are two bounds on this evolution. First, for the specific case of time evolution ($\theta=t$) it was shown by Mandelstam and Tamm that a lower bound could be defined in terms of the {\em variance in the energy} of the system \cite{mandelstam45}:
\be \label{eq:MT1}
t \geq \frac{\pi}{2} \frac{\hbar}{\Delta E}\, ,
\ee
with $(\Delta E)^2 = \la H^2 \ra - \la H \ra^2$, and $H$ is the Hamiltonian of the system. Relatively recently, Margolus and Levitin derived a second bound on the the minimum time of unitary evolution, which is instead determined by the {\em average energy} of the system \cite{margolus98}:
\be \label{eq:ML1}
t \geq \frac{\pi}{2} \frac{\hbar}{E}\, ,
\ee
where $E = \la H \ra$. For the purposes of determining the so-called maximum speed of evolution for a quantum system, having a bound defined in terms of the average energy of the system is in general far more useful than a bound in terms of the variance, since the average energy is generally easier to determine than the variance in the energy. For example, the Margolus-Levitin bound was used to estimate the ultimate limits to computation and the maximum computational power of the universe \cite{lloyd00,lloyd02}.

In this paper, we give an alternative derivation of the Mandelstam-Tamm and Margolus-Levitin inequalities based on the distance of quantum states in Hilbert space. In section \ref{sec:sd} we give a brief review of the concept of statistical distance, both for classical probability distributions and for quantum states, and we relate the statistical distance to the Fisher information, which measures the amount of information about a parameter obtained in a given measurement. In section \ref{sec:ineq} we define the speed of dynamical evolution as the rate of change of the statistical distance, analogous to the velocity in real space, and we use this to derive the two inequalities. Finally, we will show by means of a counterexample that the inequalities apply to unitary evolutions, and can be violated in general completely positive trace preserving maps (CP maps).

\section{Statistical distance}\label{sec:sd}

\noindent
In order to quantify the difference between two probability distributions, we can define a distance measure in the space of probability distributions, called the {\em statistical distance}. In general, a distance $s$ between two points $a$ and $b$ in a metric space has the following properties:
\begin{enumerate}\itemsep=0pt
 \item $s(a,b) \geq 0$,
 \item $s(a,b) = 0 ~\Leftrightarrow~ a=b$,
 \item $s(a,b) = s(b,a)$,
 \item $s(a,c) \leq s(a,b) + s(b,c)\quad$ (the triangle inequality).
\end{enumerate}
When the two points $a$ and $b$ are very close together ($b=a+da$), we can define the infinitesimal distance between the points in terms of a metric tensor $g_{jk}$, according to
\be\label{eq:infdist}
 ds^2 = \sum_{j,k} g_{jk} da^j da^k\, ,
\ee
where the $da^j$ are the components of the tangent vector to $a$. We make the distinction between upper and lower indices, with the upper indicating the tangent (contravariant) vector, and the lower index indicating the one-form (covariant vector). In the case of Euclidean spaces the metric will be proportional to the Kronecker delta $\delta_{jk}$, and Eq.~(\ref{eq:infdist}) reduces to Pythagoras' theorem.

\subsection{Statistical distance for classical probability distributions}

\noindent
In general, one can define many different distance measures on a metric space. As a trivial example, we can multiply any distance function by an arbitrary constant and still satisfy the four properties above. This corresponds to changing the units of length for the distance function. In our case, we want to choose a natural distance function for the space of probability distributions that relates the statistical distance to expectation values. The natural inner product in the space of probability distributions is the expectation value of a random variable $A$
\be
 \la A\ra = \sum_j A_j p^j\, ,
\ee
and the correlation is
\be\label{eq:corr}
 \la A B\ra = \sum_j A_j B_j p^j = \sum_{jk} A_j B_k g^{jk}\, .
\ee
We related the correlation to the statistical distance by writing $\la AB\ra$ in terms of a metric $g^{jk}$. Since the probability distributions form the vectors in the space, the probabilities $p^j$ are the proper tangent vectors, while the one-forms $A_j$ and $B_k$ are the values of the random variables $A$ and $B$. It is clear that the correlation forms the natural quadratic form for the space of probability distributions.

From Eq.~(\ref{eq:corr}), we find that the metric $g^{jk}$ must be proportional to $p^j$, together with a Kronecker delta $\delta_{jk}$ to match the indices. However, the metric we require for the infinitesimal form of the statistical distance in Eq.~(\ref{eq:infdist}) has lower indices. We therefore find that
\be
 g_{jk} = (g^{jk})^{-1} = \frac{\delta_{jk}}{p^j}\, .
\ee
Consequently, the distance between two probability distributions $p^j$ and $p^j + dp^j$ using this metric becomes
\be
 ds^2 = \sum_{j,k} g_{jk} dp^j dp^k = \sum_j \frac{(dp^j)^2}{p^j}\, .
\ee
This is the infinitesimal statistical distance for classical probability distributions. The distance increases without bound when one of the $p^j$s becomes zero. This is interpreted as follows: when we try to distinguish between two probability distributions $p_A$ and $p_B$, and one type of event has zero probability in $p_A$ but not in $p_B$, then the occurrence of that event immediately tells us with certainty that our system is described by $p_B$. In practical numerical applications, we exclude the boundary of the probability simplex to avoid these divergency issues.

At this point, we note a peculiarity in the form of the statistical distance for classical probability distributions. If we write the probability $p^j = (r^j)^2$ in terms of (real) probability amplitudes $r^j$, we obtain $dp^j = 2 r^j dr^j$, and the statistical distance becomes
\be\label{eq:sd}
 ds^2 = 4 \sum_j (dr^j)^2\, .
\ee
In other words, the statistical distance in the space of probability distributions is the Euclidean distance in the space of (real) probability amplitudes. However, probability amplitudes are usually associated with quantum mechanics, and our entire discussion has been classical. Note also the factor 4 in the statistical distance. We could remove it by rescaling the distance units. However, our units are quite convenient, and the factor will become important in section \ref{sec:ineq}, when we calculate the speed of dynamical evolution.

So far, we have been diligent in observing the difference between upper and lower indices in order to derive the correct form of the metric tensor. In the remainder of this paper, this is no longer necessary, and we will use only lower indices from now on.

\subsection{Statistical distance for density matrices}

\noindent
Next, we will derive the statistical distance for density matrices $\rho$. We will follow the general procedure of the previous section, and derive the equivalent of the metric tensor by identifying the natural quadratic form. Again, the natural inner product on the space of density matrices is the Born rule for the expectation value of a quantum mechanical observable $A$, given by
\be
 \la A \ra = \tr(\rho A)\, .
\ee
Similarly, we can choose the correlation as the natural quadratic form. However, there is a subtlety: since correlations are observable, they must be represented by Hermitian operators. On the other hand, when we consider the correlation between two non-commuting observables the operator product $AB$ is not Hermitian: $(AB)^{\dagger} = B^{\dagger} A^{\dagger} = BA \neq AB$. The correct correlation is therefore given by the symmetrized product of $A$ and $B$:
\be
 \frac{1}{2}\la AB+BA \ra = \frac12 \tr[\rho\{A,B\}]\, ,
\ee
where we used the anti-commutator $\{A,B\} = AB+BA$. Using the cyclic property of the trace, this can be rewritten as
\be
  \frac{1}{2}\la AB+BA \ra = \tr\left[A\mathcal{R}_{\rho}(B)\right] ,
\ee
with
\be
 \mathcal{R}_{\rho}(B) = \frac{1}{2} \{\rho,B\} = \frac{1}{2}\sum_{j,k} (p_j + p_k) B_{jk} \ket{j}\bra{k}\, ,
\ee
and its inverse
\be
 \mathcal{L}_{\rho}(B) = \mathcal{R}^{-1}_{\rho}(B) = \sum_{j,k} \frac{2B_{jk}}{p_j + p_k}  \ket{j}\bra{k}\, .
\ee
Here, we used the diagonal basis for $\rho = \sum_j p_j \ket{j}\bra{j}$. The ``raising'' and ``lowering'' operators $\mathcal{R}_{\rho}$ and $\mathcal{L}_{\rho}$ play the role of the metric, and the infinitesimal statistical distance between $\rho$ and $\rho+d\rho$ on the space of density operators can be written as \cite{braunstein94}
\be\label{eq:qsd}
 ds^2_{\rho} = \tr\left[ d\rho \mathcal{L}_{\rho}(d\rho) \right]\, .
\ee
In the case of pure quantum states this distance simplifies considerably, and we obtain the Wootters distance \cite{wootters81}. If $\ket{\psi}$ and $\ket{\phi}$ are two arbitrary pure quantum states in Hilbert space, then the statistical distance is given by the angle between the two states:
\be \label{eq:Wootters}
 s(\psi,\phi) = \arccos|\la\psi|\phi\ra|\, .
\ee
Notice how the Wootters distance is related to the fidelity $|\la\psi|\phi\ra|^2$, which is the probability of mistaking $\ket{\psi}$ for $\ket{\phi}$ in a single-shot measurement.

\subsection{Relation to the Fisher Information}

\noindent
The statistical distance is a measure of how close one probability distance is to another. In other words, it can be directly related to the number of times we need to sample our system in order to tell reliably which of the two probability distributions describes our system. Imagine that the two probability distributions are connected by a curve that is parametrized by a real number $\theta$. Distinguishing between the two distributions then reduces to the estimation of the parameter $\theta$. This is a well-known problem, and the information about $\theta$ in a particular measurement procedure is given by the Fisher information $F(\theta)$. For a discrete set of possibilities, the Fisher information is
\be\label{eq:fisher}
 F(\theta) = \sum_j \frac{1}{p_j(\theta)} \left( \frac{dp_j(\theta)}{d\theta} \right)^2 = \left( \frac{ds}{d\theta} \right)^2\, ,
\ee
where $p_j(\theta)$ is determined by the Born rule
\be
 p_j(\theta) = \tr[E_j \rho(\theta)]\, ,
\ee
with $E_j$ the POVM associated with measurement outcome $j$. A similar expression holds when the possible events form a continuum.

We see from Eq.~(\ref{eq:fisher}) that the Fisher information is directly related to the derivative of the statistical distance, as expected. Moreover, using the expression in Eq.~(\ref{eq:qsd}) for the quantum mechanical statistical distance, we obtain
\be\label{eq:fm}
 F(\theta) = \left( \frac{ds}{d\theta} \right)^2 = \tr\left[ \rho' \mathcal{L}_{\rho}(\rho') \right] ,
\ee
where $\rho'$ is the derivative of $\rho$ with respect to $\theta$. If translations in $\theta$ are generated by a Hermitian operator ${K}$, we can use the Heisenberg equation of motion to write
\be
 \frac{d\rho}{d\theta} = \frac{1}{i\hbar} [K,\rho] = \frac{1}{i\hbar} [K - \la K\ra,\rho] = \frac{1}{i\hbar} [\Delta K,\rho] \, ,
\ee
where $\la K\ra = \tr(\rho K)$ is a real number, and can therefore be included in the commutator with impunity. Substituting this into Eq.~(\ref{eq:fm}) yields \cite{braunstein96b}
\ba\label{eq:fb}
 F(\theta) &=& \tr\left[ \rho' \mathcal{L}_{\rho}(\rho') \right] = \frac{2}{\hbar^2}\sum_{j,k} \frac{(p_j - p_k)^2}{p_j+p_k} |\Delta K_{jk}|^2 \cr &\leq& \frac{4}{\hbar^2} \left\la (\Delta K)^2\right\ra\, .
\ea
In other words, the amount of information about $\theta$ in any measurement is bounded by the variance in its generator $K$ \cite{braunstein96b}.

% we cannot use this trick to derive the Margolus-Levitin bound, because the square is inside the expectation value, and we can never get the average energy from this.

\section{Dynamical evolution of quantum states}\label{sec:ineq}

\noindent
The quantum speed limit for the dynamical evolution of quantum states will take the form of two inequalities, the Mandelstam-Tamm inequality and the Margolus-Levitin inequality. Before we derive these inequalities, we will have to define precisely what we mean by the speed of dynamical evolution. Analogous to the instantaneous velocity of a particle in real space, which is the time derivative of the position function of the particle, we define the dynamical speed of evolution for quantum states as the derivative of the statistical distance function with respect to the parameter $\theta$:
\be
 v(\theta) = \frac{ds}{d\theta}\, .
\ee
This allows us to use the results from the previous section to obtain bounds on $v(\theta)$.

\subsection{Mandelstam-Tamm inequality}

\noindent
The Mandelstam-Tamm inequality is almost immediate from Eq.~(\ref{eq:fb}):
\be
 F(\theta) = \left( \frac{ds}{d\theta} \right)^2 \leq \frac{4}{\hbar^2} \left\la (\Delta K)^2\right\ra\, .
\ee
Taking the positive roots, we can rewrite this as
\be
 \frac{ds}{d\theta} \leq \frac{2}{\hbar} \delta K\, ,
\ee
where $\delta K \equiv |\sqrt{\la(\Delta K)^2\ra}|$. Separating the variables and integrating yields
\be\label{eq:mtm1}
 \int_0^{\theta} d\theta' \geq \frac{1}{2} \frac{\hbar}{\delta K} \int_0^{\pi} ds \quad\Rightarrow\quad
 \theta \geq \frac{\pi}{2} \frac{\hbar}{\delta K}\, .
\ee
In the case where $\theta$ is the time parameter generated by the Hamiltonian $H$, the inequality reduces to the famous Mandelstam-Tamm inequality
\be\label{eq:mt2}
 t \geq \frac{\pi}{2}\frac{\hbar}{\Delta E}\, .
\ee
Note that the integral over $ds$ in Eq.~(\ref{eq:mtm1}) runs from 0 to $\pi$, instead of $\pi/2$. This is due to the factor 4 in Eq.~(\ref{eq:sd}). Alternatively, this can be seen in the Bloch sphere, were the angle between orthogonal qubit states is $\pi$, rather than $\pi/2$. Note that the Mandelstam-Tamm inequality in Eq.~(\ref{eq:mt2}) was derived for unitary evolution of {\em arbitrary mixed states}. We will see in Section \ref{sec:counter} that this bound can be violated by non-unitary evolutions of density operators.

\subsection{Margolus-Levitin inequality}

\noindent
Instead of using the bound on the Fisher information directly to derive the Mandelstam-Tamm inequality, we can use the expression for the statistical distance and carry out the differentiation with respect to $\theta$ explicitly. Remarkably, this will yield the Margolus-Levitin inequality. We will proceed by first deriving the inequality for pure states, followed by an extension to mixed states via a standard purification procedure.

We again consider the evolution parameterized by $\theta$, which is generated by the Hermitian operator ${K}$. In this case, a system initially described by the pure state $\ket{ \psi_0 }$ at $\theta = 0$ will evolve to
\be \label{eq:evo}
\ket{ \psi_\theta } = \exp\left(-\frac{i}{\hbar}{K}\theta\right)\ket{ \psi_0 }\, .
\ee
The Wootters distance between two pure states is given by the angle between the states and the rate of change of the statistical distance can therefore be written as
\begin{eqnarray} \label{eq:d1}
\frac{ds}{d\theta} &=& \frac{d}{d\theta}\arccos\left(|\la\psi_0|\psi_\theta\ra| \right)  \cr
&=& -\frac{1}{\sqrt{1-|\la\psi_0|\psi_\theta\ra|^2 }}\frac{d}{d\theta}|\la\psi_0|\psi_\theta\ra|\, .
\end{eqnarray}
Since the pre-factor $1/\sqrt{(1 - x^2)} \geq 1$ for all real $x$, we obtain the inequality
\be \label{eq:d3}
\frac{ds}{d\theta} \le - \frac{d}{d\theta} |\la\psi_0|\psi_\theta\ra|\, .
\ee
Next, we prove that
\be \label{eq:d4}
-\frac{d}{d\theta} |\la\psi_0|\psi_\theta\ra| \le \left|\frac{d}{d\theta}\la\psi_0|\psi_\theta\ra\right|\, .
\ee
To this end, we rewrite the derivative on the left-hand side of Eq.~(\ref{eq:d4}) as
\be \label{eq:d5}
\frac{d}{d\theta} |\la\psi_0|\psi_\theta\ra| = \frac{d}{d\theta}\sqrt{\la\psi_0|\psi_\theta\ra\la\psi_\theta|\psi_0\ra}\, ,
\ee
and using the generalised Schr{\"o}dinger equation
\be
 i\hbar \frac{d}{d\theta} \ket{\psi_{\theta}}={K}\ket{\psi_{\theta}}\, ,
\ee
this becomes
\begin{eqnarray} \label{eq:d6}
\frac{d}{d\theta} |\la\psi_0|\psi_\theta\ra| &=& \frac{-i\bra{\psi_0}{K}\ket{\psi_\theta}\la\psi_\theta|\psi_0\ra+i\la\psi_0|\psi_\theta\ra\bra{\psi_\theta}{K}\ket{\psi_0}}{2\hbar|\la\psi_0|\psi_\theta\ra|} \nonumber \\
&=& \frac{\mathrm{Im} (\bra{\psi_0} K \ket{\psi_\theta}\la\psi_\theta|\psi_0\ra)}{\hbar|\la\psi_0|\psi_\theta\ra|} \nonumber \\
&\le& \frac{|\bra{\psi_0}{K}\ket{\psi_\theta}\la\psi_\theta|\psi_0\ra|}{\hbar|\la\psi_0|\psi_\theta\ra|}.
\end{eqnarray}
The right-hand side of Eq.~(\ref{eq:d4}) becomes
\begin{eqnarray} \label{eq:d7}
\left|\frac{d}{d\theta}\la\psi_0|\psi_\theta\ra\right|&=& \frac{1}{\hbar}|\bra{\psi_0}{K}\ket{\psi_\theta}| = \frac{|\bra{\psi_0}{K}\ket{\psi_\theta}|.|\la\psi_0|\psi_\theta\ra|}{\hbar|\la\psi_0|\psi_\theta\ra|} \nonumber \\
&\ge& \frac{|\bra{\psi_0}{K}\ket{\psi_\theta}\la\psi_0|\psi_\theta\ra|}{\hbar|\la\psi_0|\psi_\theta\ra|}  \, ,
\end{eqnarray}
where we again used the generalized Schr{\"o}dinger equation, and in the last line we used the Cauchy-Schwarz inequality. Finally, we combine Eq.~(\ref{eq:d7}) and Eq.~(\ref{eq:d6}) to obtain
\begin{equation} \label{eq:d8}
\frac{d}{d\theta} |\la\psi_0|\psi_\theta\ra| \le \frac{\mathrm{Im}(\bra{\psi_0}{K}\ket{\psi_\theta}\la\psi_\theta|\psi_0\ra)}{\hbar|\la\psi_0|\psi_\theta\ra|} \le  \left|\frac{d}{d\theta} \la\psi_0|\psi_\theta\ra\right|.
\end{equation}
Since $\arccos(x)$ is a monotonically decreasing function in the interval $0\leq x \leq 1$, the derivative of $|\la\psi_{0}\ket{\psi_{\theta}}\ra|$ with respect to $\theta$ is strictly positive and we have therefore proved Eq.~(\ref{eq:d4}).

Continuing the derivation of the Margolus-Levitin inequality, we use Eq.~(\ref{eq:d8}) in Eq.~(\ref{eq:d3}) and find that
\begin{equation} \label{eq:d9}
\frac{ds}{d\theta} \leq \left|\frac{d}{d\theta}\la\psi_0|\psi_\theta\ra \right| \leq \frac{|\bra{\psi_0}{K}\ket{\psi_\theta}|}{\hbar} \leq \frac{|\bra{\psi_0}{K}\ket{\psi_0}|}{\hbar} \equiv \frac{|\la{K}\ra|}{\hbar} .
\end{equation}
Separating the variables $s$ and $\theta$, we obtain
\be \label{eq:d10}
\int_0^\theta  d\theta' \ge \frac{\hbar}{|\la{K}\ra|} \int_0^\frac{\pi}{2} ds\, ,
\ee
and integrating both sides gives
\be \label{eq:d11}
\theta \ge  \frac{\pi}{2}\frac{\hbar}{|\la{K}\ra|}\, .
\ee
In the case where ${K}$ is the Hamiltonian and $\theta$ the time, the inequality becomes the Margolus-Levitin inequality
\be \label{eq:ML}
t \ge  \frac{\pi}{2}\frac{\hbar}{E}\, ,
\ee
with $E$ the average energy of the system.

So far, we have shown that the Margolus-Levitin inequality holds for the unitary evolution of pure states. However, following the derivation of the Mandelstam-Tamm inequality, we would like to extend this bound to the unitary evolution of arbitrary mixed states. The challenge is to find the most convenient distance measure between mixed states, i.e., a generalization of the Wootters distance for density operators. We choose the purification of the density matrices that maximizes the fidelity between them. The Margolus-Levitin inequality then applies to the purifications, which in turn can be translated to a bound on the speed of unitary evolution of mixed states.

The fidelity between two density matrices $\rho$ and $\sigma$ is defined as \cite{jozsa94}
\be \label{eq:fd}
 F(\rho,\sigma) \equiv \left[ \tr\left(\sqrt{\rho^{\frac12}\sigma\rho^{\frac12}}\right) \right]^2\, ,
\ee
which can be interpreted as  the probability of mistaking $\rho$ for $\sigma$ in a single-shot measurement. Despite its appearance, $F(\rho,\sigma)$ is symmetric in $\rho$ and $\sigma$. We can relate the fidelity to a distance measure in various ways, one possibility being
\be
 s(\rho,\sigma) = \arccos\left[\sqrt{F(\rho,\sigma)}\right]\, .
\ee
When both $\rho$ and $\sigma$ are pure states $s(\rho,\sigma)$ reduces to the Wootters distance of Eq.~(\ref{eq:Wootters}). According to Uhlmann's theorem \cite{uhlmann76}, the square root of the fidelity is given by
\be\label{eq:UT}
  \sqrt{F(\rho,\sigma)} = \max_{\ket{\chi_\rho},\ket{\chi_\sigma}}|\la\chi_\rho|\chi_\sigma\ra|\, ,
\ee
where $\ket{\chi_\rho}$ and $\ket{\chi_\sigma}$ are purifications of $\rho$ and $\sigma$. In other words, if $\rho$ is the state of a system $S$, and $\ket{\chi_{\rho}}$ is a pure state of a compound system $S+R$, then $\rho = \tr_R(\ket{\chi_{\rho}}\bra{\chi_{\rho}})$. The system $R$ must be described on a Hilbert space that is at least as large as the Hilbert space of system $S$. A similar definition holds for $\ket{\chi_{\sigma}}$. For our purposes it is sufficient to note that there exist states $\ket{\chi_\rho}$ and $\ket{\chi_\sigma}$ for which the equality holds: $F(\rho,\sigma) = |\la \chi_{\rho}|\chi_{\sigma}\ra|^2$.

For pure states the Wootters distance is interpreted as the angle between the states in Hilbert space. For mixed states, this interpretation makes sense only when the states have the same purity $\tr(\rho^2) = \tr(\sigma^2)$, and the evolution is unitary. The speed of evolution can then be bounded using
\begin{equation}\label{eq:MixedRate}
\frac{ds}{d\theta} =
\frac{d}{d\theta}\arccos\left[\sqrt{F(\rho_0,\rho_\theta)}\right] = \frac{d}{d\theta}\arccos|\la\chi_0|\chi_{\theta}\ra|\, ,
\end{equation}
where $\rho_0$ is the initial state, and $\rho_{\theta}$ is the evolved state. The pure states $\ket{\chi_0}$ and $\ket{\chi_\theta}$ are the respective purifications that maximize the fidelity. Since $\ket{\chi_0}$ and $\ket{\chi_{\theta}}$ are pure, this is just the derivative of the standard Wootters distance in the compound system $S+R$, and the same argument following Eq.~(\ref{eq:d1}) holds. We therefore recover the Margolus-Levitin inequality for the purified compound system:
\be \label{eq:MLPure} \theta_{p}
 \geq \frac{\pi}{2}\frac{\hbar}{\la K_S\ra + \la K_R\ra} = \frac{\pi}{2}\frac{\hbar}{\la K\ra} \, ,
\ee
where $\la K_S\ra$ is the expectation value of $K$ on the system $S$, and $\la K_R\ra$ the expectation value on $R$. In order to interpret this inequality, let $K$ be the Hamiltonian of the system. The expectation value $\la K_S\ra + \la K_R\ra$ is then the average energy of the system $S$ and the purification system $R$ taken together. At first glance this seems problematic since the purification is merely a mathematical construction and therefore unrestricted amounts of energy may be added to the system during the process. This would lead to arbitrarily short orthogonality times. However it is precisely because the purification is a mathematical construct that this is not a problem: any extra energy added is physically meaningless and therefore the true bound on the evolution occurs for $\la H\ra = \la H_S\ra$. In other words we simply have to choose purifications made up only of degenerate ground eigenstates (with $E=0$). This means that the evolution of mixed states is also bounded by Eq.~(\ref{eq:ML}). A similar line of reasoning holds for general Hermitian operators $K$. It was show in Ref.~\cite{levitin09} that mixed states can never attain either equality. In addition, any state that does attain the inequalities has the form \cite{levitin09, soderholm99}
\be \label{eq:puresat}
\ket{\psi} = \frac{1}{\sqrt{2}}(\ket{\psi_{0}}+e^{i\theta}\ket{\psi_{n}})\, ,
\ee
where the states $\ket{\psi_{n}}$ are (possibly degenerate) energy eigenstates of the system.

\section{Non-unitary evolution}\label{sec:counter}

\begin{figure}[t]
 \includegraphics[width=8.5cm]{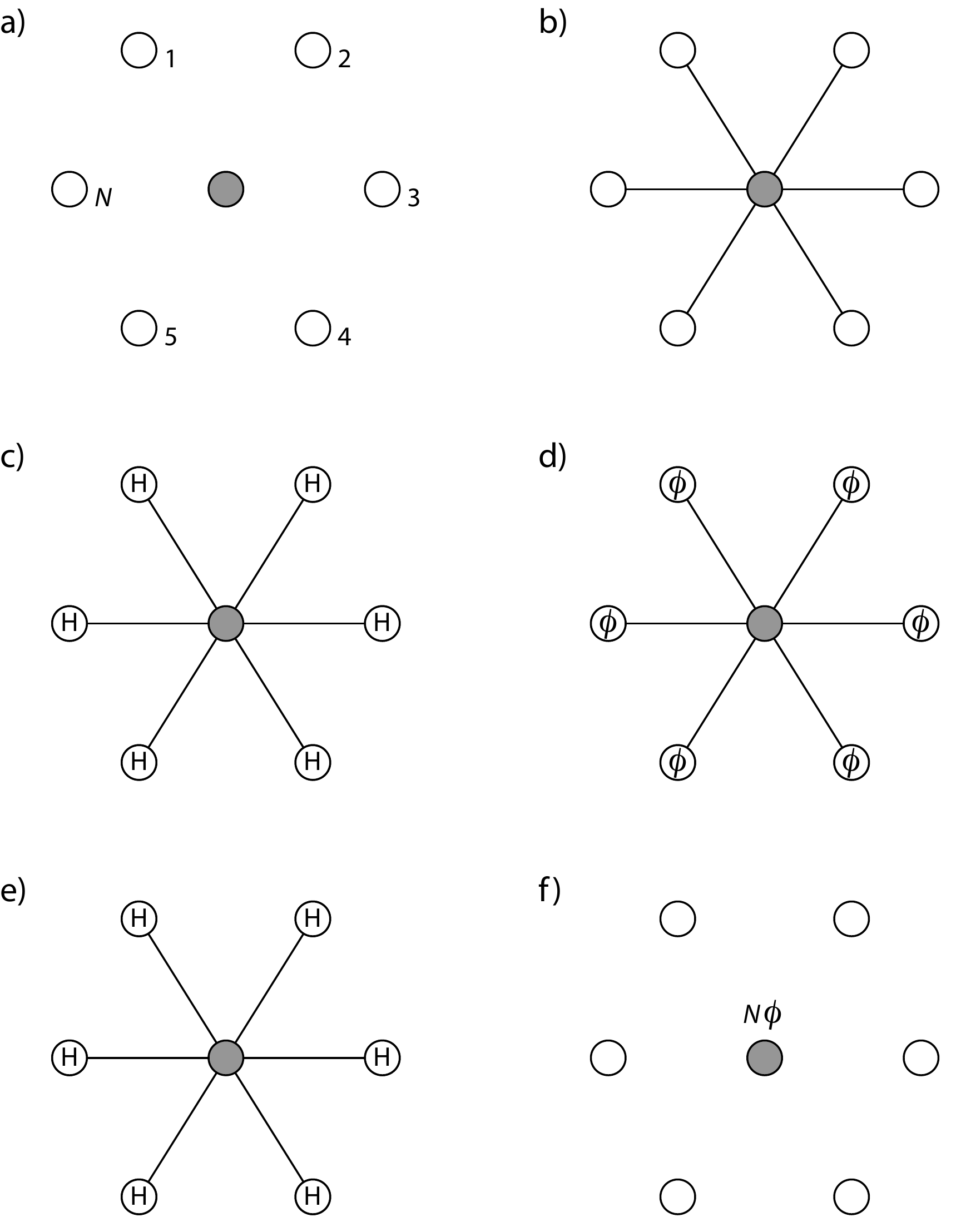}
 \caption{Non-unitary evolution of the central qubit that violates both the Mandelstam-Tamm and Margolus-Levitin bound: (a) All qubits are prepared in the state $\ket{+}$; (b) each qubit is entangled simultaneously with the central qubit using a $CZ$ gate; (c) Hadamard gates are applied to the satellite qubits, which creates a GHZ state of size $N+1$; (d) phase shifts $\phi$ are applied to the satellite qubits, and (e) a second set of Hadamard gates are applied to the satellite qubits; (f) $N$ simultaneous $CZ$ gates disentangle the satellite qubits from the central qubit, leaving it with an accumulated phase shift of $N\phi$.}\label{fig:counterexample}
\end{figure}

\noindent
The Mandelstam-Tamm and Margolus-Levitin inequalities do not hold for the most general quantum evolutions, described by completely positive, trace preserving maps (CP maps). To demonstrate this, we construct a counterexample. We can always describe a non-unitary evolution of a single system as the unitary evolution of the system and its environment combined. The non-unitary evolution is obtained when we trace out the environment of the system. In our counterexample, the system starts and ends in a separable state with respect to the environment, which allows us to compare the states before and after the interaction. We show that orthogonal initial and final states can be created in arbitrary short times.

Consider a two-level system (a qubit) with an energy gap $E_0$ between the ground state $\ket{0}$ and the excited state $\ket{1}$. When the qubit is prepared in the eigenstate of the Pauli $X$ operator, the minimum time for this system to evolve unitarily to an orthogonal state is
\be\label{eq:timee0}
 t = \frac{\pi\hbar}{E_0}\, ,
\ee
saturating both bounds, with $\Delta E = \la H\ra = E_0/2$. In addition, consider $N$ qubits of a possibly different species with energy gap $E_q$. Our setup is shown in Fig.~\ref{fig:counterexample}. We call the qubit with energy gap $E_0$ the `central qubit', and the remaining qubits are the `satellite qubits'. At time $t=0$, all qubits are prepared in the state $\ket{+} = (\ket{0}+\ket{1})/\sqrt{2}$. To describe the counterexample we use the stabilizer formalism, in which the state of a system of qubits is fully determined as the $+1$ eigenvalue state of a set of operators $S_i$, called the stabilizer generators that generate an abelian group. For a detailed exposition of the stabilizer formalism, see Nielsen and Chuang \cite{nielsen00}. The initial state of the central qubit (labelled `0') is then determined fully by the operator $\smash{S_0^{\rm(a)} = X_0}$. The satellite qubit $j$ is stabilized by $\smash{S_j^{\rm(a)}} = X_j$, with $j=1,\ldots,N$.

Next, we entangle the qubits using $CZ$ gates, causing the stabilizer generators to evolve to
\be \label{eq:Stable}
 S_0^{\rm(a)} \to S_0^{\rm(b)} = X_0 \prod_{j=1}^N Z_j \quad\text{and}\quad S_j^{\rm(a)} \to S_j^{\rm(b)} = Z_0 X_j\, .
\ee
This evolution occurs in time $\smash{t_{CZ}^{(N)}}$. The subsequent Hadamard gate on the satellite qubits changes the stabilizer generators to
\be
 S_0^{\rm(c)} = \prod_{j=0}^N X_j \quad\text{and}\quad S_j^{\rm(c)} = Z_0 Z_j\, .
\ee
This evolution will take a time $\smash{t_H^{(N)}}$. It is straightforward to check that the state state corresponding to these stabilizer generators is the GHZ state $(\ket{0,\ldots,0}+\ket{1,\ldots,1})/\sqrt{2}$ on the $N+1$ qubits. The phase shift $\phi$ on the satellite qubits given by $U_j(\phi) = \exp(i\phi Z/2)$ change the stabilizer generators to
\be
 S_0^{\rm(d)} = X_0 \prod_{j=1}^N U_j(\phi) X_j U_j^{\dagger}(\phi) = X_0 \prod_{j=1}^N (\cos\phi\, X_j + \sin\phi\, Y_j)\, ,
\ee
and the stabilizer generators for the satellite qubits remain unchanged: $\smash{S_j^{\rm(d)} = S_j^{\rm(c)}}$. This evolution will take a time $\smash{t_{\phi}^{(N)}}$. After the phase evolution, we again apply Hadamard gates to the satellite qubits, leading to the stabilizer generators
\be
 S_0^{\rm(e)} = X_0 \prod_{j=1}^N (\cos\phi\, Z_j - \sin\phi\, Y_j) \quad\text{and}\quad S_j^{\rm(e)} = Z_0 X_j\, .
\ee
Finally, the satellite qubits are disentangled from the central qubit with another $N$ $CZ$ gates, leading to
\be
 S_0^{\rm(f)} = \cos(N\phi)\, X_0 + \sin(N\phi)\, Y_0 \quad\text{and}\quad S_j^{\rm(f)} = X_j\, .
\ee
In other words, the satellite qubits are back to their initial state, and the central qubit is in the state
\be
 \ket{\psi^{\rm(f)}} = \frac{\ket{0} + e^{iN\phi} \ket{1}}{\sqrt{2}}\, .
\ee
This is orthogonal to the initial state of the central qubit when $N\phi = \pi$. The total time $\tau$ taken by this evolution is
\be
 \tau = 2t_{CZ}^{(N)} + 2 t_H^{(N)} + t_{\phi}^{(N)}\, .
\ee
In order to show that this is a counterexample to the Mandelstam-Tamm and Margolus-Levitin bounds, we need to show that we can choose parameters such that $\tau < \pi\hbar/E_0$.

The Hadamard gate evolves a state halfway to an orthogonal state, and the speed limit for this gate can therefore be taken as half the time given in Eq.~(\ref{eq:timee0}), with $E_0 \to E_q$:
\be
 t_H = \frac{\pi\hbar}{2 E_q}\, .
\ee
Since all Hadamard gates are applied simultaneously to different qubits, the minimum time to apply $N$ Hadamard gates is the same as that for a single Hadamard gate. Therefore, we have $\smash{t_H^{(N)} = t_H}$. A similar argument holds for the application of the phase shifts on the satellite qubits, leading to $\smash{t_{\phi}^{(N)} = t_{\phi}}$.

The determination of the time it takes to apply the $N$ $CZ$ gates requires a little more care, since each gate involves the central qubit. First, consider the time evolution of $S_{j}$
\be \label{eq:HE}
\frac{dS_{j}}{dt}=\frac{i}{h}[H,S_{j}].
\ee
The interaction Hamiltonian for a $CZ$ gate between the central qubit and the $j^{\rm th}$ satellite qubit can be written as $H_j=  g(\unity_{0}-Z_{0})(\unity_{j}-Z_{j})$. Since $[H_j,H_k]=0$ for all $j$ and $k$, the interaction Hamiltonian for the $N$ $CZ$ gates is $H=\sum_j H_j$.  Solving the Heisenberg equation in Eq.~(\ref{eq:HE}) gives
\begin{eqnarray} \label{eq:S(t)}
S_{j}(t)&=&\frac{1}{2}\left[1+\cos\left(\frac{gt}{\hbar}\right)\right]X_{j}+\frac{1}{2}\sin\left(\frac{gt}{\hbar}\right)Y_{j}+ \nonumber \\ && \frac{1}{2}\left[1-\cos\left(\frac{gt}{\hbar}\right)\right]Z_{0}Y_{j}-\frac{1}{2}\sin\left(\frac{gt}{\hbar}\right)Z_{0}Y_{j}\, ,
\end{eqnarray}
which exhibits a periodic behaviour with period $T=2\pi\hbar/g$. After half this period, the stabilizer $S_j(t)$ becomes $\smash{S_j^{(\rm b)}}$, and the gate time for a single $CZ$ gate is therefore 
\be
 t_{CZ} = \frac{1}{2} T = \frac{\pi \hbar}{g}\, .
\ee
In the case of $N$ $CZ$ gates, the evolution of the stabilizer generator $S_0(t)$ is more complicated, since it encapsulates the interaction of the central qubit with all the satellite qubits. We have to show that it nevertheless exhibits the same periodicity as $S_j(t)$. In general, we can write $S_0(t)$ in terms of the Pauli operators on the central qubit and the $j^{\rm th}$ satellite qubit, while collecting all other Pauli operators in $Q^{(j)}$:
\begin{eqnarray} \label{eq:S0}
S_{0}(t) &=& \alpha_j(t) Q^{(j)}_{\alpha}\, X_0 \unity_j + \beta_j (t)Q^{(j)}_{\beta}\, Y_0 \unity_j\cr && + \gamma_j (t) Q^{(j)}_{\gamma}\, X_0 Z_j +\delta_j (t)Z_{0} Q^{(j)}_{\delta}\, Y_0 Z_j\, ,
\end{eqnarray}
where $\alpha_j (t)$, $\beta_j (t)$, $\gamma_j (t)$, and $\delta_j (t)$ are the coefficients that determine the time behaviour. Their exact form is not important for our argument. Since the operators $S_0(t)$ and $S_j(t)$ stabilize the evolving state at all times, and the stabilizer is an abelian group, we require that $[S_0(t),S_j(t)]=0$ for all $t$. The periodicity of $S_j(t)$ then implies the periodicity of $S_0(t)$, and moreover, they must have the same period. Since this period is independent of $N$, we arrive at the conclusion that all $N$ $CZ$ gates can be applied simultaneously. Notice that although the coefficients of $S_0(t)$ will change for larger $N$,
the underlying periodicity cannot be affected without causing the commutation relations $[S_0(t),S_j(t)]$ to become nonzero.

Finally, we choose $t_{\phi}$ (and therefore $\phi$ itself) arbitrarily small, and $N$ arbitrarily large, such that $N\phi=\pi$ and $t_{\phi} \approx 0$. The parameter regime where the system becomes a counterexample to the quantum speed limits ($\tau < \pi\hbar/E_0$) then becomes
\be
 \frac{2\pi\hbar}{g} + \frac{\pi\hbar}{E_q} <  \frac{\pi\hbar}{E_0}\, .
\ee
This can be re-cast into the following two requirements:
\be
 E_q > E_0 \quad\text{and}\quad g > \frac{2E_q E_0}{E_q - E_0}\, .
\ee
Since this is a perfectly valid parameter regime in quantum mechanics, this constitutes a proper counterexample (even though it may be very difficult to implement in practice). It demonstrates that the bounds are valid only for unitary evolution. In fact, $\tau$ can be made arbitrarily small by making $E_q$ and $g$ arbitrarily large, and consequently there is no general bound that is valid for all possible (unitary and non-unitary) quantum evolution.

The violation of the Mandelstam-Tamm and Margolus-Levitin bounds leads to a number of open questions. First, what are the general requirements on a system and its interaction with the environment in order to violate the speed limit? The counterexample presented here suggests that the interaction between the system and its environment must be strong, and that the environment must be more energetic than the system ($E_q > E_0$). Second, can these regimes be probed with experiments? It is not clear {\em a priori} that the strong interaction of our counterexample can be achieved in a noiseless way. This may prevent the practical implementation of the evolution, even though the counterexample is perfectly valid from a mathematical point of view. In other words, taking into account noise may lead to another practical speed limit after all. And third, are there general bounds on the speed of quantum evolution based on general properties of the environment, rather than on detailed knowledge of the dynamics of the combined system? These questions must be addressed in order to fully understand the speed of dynamical evolution of open quantum systems.

\section{Conclusions}

\noindent
The Mandelstam-Tamm and Margolus-Levitin inequalities play an important role in the study of quantum mechanical processes in Nature. However, to date there has been only one derivation of the Margolus-Levitin inequality. In this paper, we gave alternative derivations for both inequalities from the statistical distance between quantum states. This allows for a fully geometrical interpretation of the quantum speed limit. The inequalities were shown to hold for unitary evolution of pure and mixed states, and a counterexample to the inequalities is given for evolution described by completely positive trace-preserving maps.

The authors thank Carlos P\'erez-Delgado, David Whittaker, and Samuel Braunstein for valuable discussions.

\end{document}